\def\beq#1{\begin{equation}\label{#1}}
\def\eeq{\end{equation}}
\def\beqa#1{\begin{eqnarray}\label{#1}}
\def\eeqa{\end{eqnarray}}

\def\fun#1#2{\lower3.6pt\vbox{\baselineskip0pt\lineskip.9pt
        \ialign{$\mathsurround=0pt#1\hfill##\hfil$\crcr#2\crcr\sim\crcr}}}

\def\xi{{{\bf x}^b}}

\documentclass
[twocolumn,aps,showpacs,showkeys,nofootinbib]{revtex4}
\usepackage{epsfig}

\newcommand{\be}{\begin{equation}}
\newcommand{\ee}{\end{equation}}
\newcommand{\ba}{\begin{eqnarray}}
\newcommand{\ea}{\end{eqnarray}}

\begin{document}
\input{epsf.sty}

\title{Exploring the Systematic Uncertainties of Type Ia Supernovae\\
 as Cosmological Probes}

\author{Shuang Wang$^{1,2}$\footnote{swang@ou.edu} and Yun Wang$^1$\footnote{wang@nhn.ou.edu}}
\address{$^1$Homer L. Dodge Department of Physics \& Astronomy, Univ. of Oklahoma, 440 W Brooks St., Norman, OK 73019, U.S.A.\\
$^2$Department of Physics, College of Sciences, Northeastern University, Shenyang 110004, China}

\begin{abstract}

We explore the systematic uncertainties of using Type Ia supernovae (SNe Ia) as cosmological probes,
using the Supernova Legacy Survey Three Year data (SNLS3).
We focus on studying the possible evolution of the stretch-luminosity parameter $\alpha$ and the color-luminosity parameter $\beta$,
by allowing $\alpha$ and $\beta$ to be function of redshift, $z$. We find no evidence for the redshift evolution of $\alpha$.
We find that without flux-averaging SNe, $\beta$ is consistent with being a constant when only
statistical uncertainties are included, but it increases significantly with $z$ when systematic uncertainties
are also included. The evolution of $\beta$ becomes marginal when all the SNe are flux-averaged, and $\beta$
is consistent with being a constant when only SNe at $z\ge 0.04$ are flux-averaged.
Our results are insensitive to the lightcurve fitter used to derive the SNLS3 sample, or the functional form
of $\alpha(z)$ and $\beta(z)$ assumed. It is likely that the apparent evolution of $\beta$ with $z$ for SNe without flux-averaging
is a consequence of unknown systematic effects; flux-averaging reduces the impact of these effects by averaging
them within each redshift bin. Assuming constant $\alpha$ and $\beta$, we find that the flux-averaging of SNe has
a significant impact on the distance-redshift relation.

\end{abstract}

\pacs{98.80.-k, 95.36.+x, 98.80.-k}

\keywords{Cosmology, Type Ia supernova}

\maketitle

\section{Introduction}

Various astronomical observations \cite{Riess98,spergel03,Tegmark04,Komatsu09,Percival10,Drinkwater10,Riess11}
all indicate that the Universe is undergoing an accelerated expansion.
So far, we are still in the dark about the nature of this extremely counterintuitive phenomenon;
it may be due to an unknown energy component (i.e., dark energy (DE) \cite{quint,phantom,k,Chaplygin,tachyonic,HDE,hessence,YMC,others1,others2,WangTegmark05,others3}),
or a modification of general relativity (i.e., modified gravity (MG) \cite{SH,PR,DGP,GB,Galileon,FR,FT,FRT}).
For recent reviews, see \cite{CST,FTH,Linder,CK,Uzan,Tsujikawa,NO,LLWW,CFPS,YWBook}.

One of the most powerful probes of DE is the use of Type Ia supernovae (SNe Ia),
which can be used as cosmological standard candles to measure the expansion history of the Universe.
In recent years, several supernova (SN) datasets with hundreds of SNe Ia were released,
such as ``Union'' \cite{Union}, ``Constitution'' \cite{Constitution}, ``SDSS'' \cite{SDSS},
``Union2'' \cite{Union2} and ``Union2.1'' \cite{Union2.1}.

In 2010, a high quality SN dataset from the first three
years of the Supernova Legacy Survey (SNLS3) was released \cite{SNLS3}.
Soon after, Conley et al. (2011; hereafter C11) presented SN-only
cosmological results by combining the SNLS3 SN with various
low- to mid-$z$ samples \cite{SNLS3Conley}, and Sullivan et al. (2011) presented
the joint cosmological constraints by combining the SNLS3 dataset with other
cosmological data sets \cite{SNLS3Sullivan}.
C11 presented three combined SN data sets, depending on
different light-curve fitters: ``SALT2'' which consists of 473 SNe Ia;
``SiFTO'', which consists of 468 SNe Ia; and ``combined'', which consists of 472 SNe Ia.
It should be stressed that, the SNLS team treated two important quantities,
stretch-luminosity parameter $\alpha$ and color-luminosity parameter $\beta$ of SNe Ia,
as free model parameters on the same footing as the cosmological parameters, all to be estimated
during the Hubble diagram fitting process using the covariance matrix that includes {\it both}
statistical and systematic errors.

A critical challenge is the control of the systematic uncertainties of SNe Ia.
One of the most important factors is the effect of potential SN evolution, i.e., the possibility of
the evolution of $\alpha$ and $\beta$ with redshift $z$.
In \cite{SDSS}, Kessler {\it et al.} showed that, for the SDSS SN sample,
there is a strong evidence for redshift-dependence of $\beta$.
In \cite{Mohlabeng}, Mohlabeng and Ralston found that $\beta$ varies with $z$ for the Union2 samples.
But in C11, the authors argued that the evolution of $\beta$ is much weaker for SNLS3 samples,
and the observed evolution may not be real (for more details, see Figure 14 of C11).

In addition, the weak lensing of high redshift SNe Ia can also lead to a systematic bias in the distances derived from SNe Ia.
In 2000, Wang pointed out that flux-averaging of SNe Ia can reduce the effect of weak lensing of SNe Ia on parameter estimation \cite{Wang2000}.
The basic idea is that because of flux conservation in gravitational lensing,
the average magnification of a large number of SNe Ia at the same redshift should be unity.
Thus averaging the observed flux from a large number of SNe Ia at the same redshift can recover the unlensed brightness of the SNe Ia at that redshift.
An interesting additional benefit of the flux-averaging of SNe Ia is that it reduces the impact of
redshift-dependent systematic biases (such as K corrections) by averaging them within each redshift bin,
see, e.g., Wang \& Tegmark (2005) \cite{WangTegmark05}.

In this paper, we explore the systematic uncertainties of SNLS3 supernova dataset, in particular,
the possible time evolution of $\alpha$ and $\beta$. We examine how this depend on the
uncertainties of data included, the lightcurve fitter used to derive the data, the functional
forms of $\alpha(z)$ and $\beta(z)$ assumed, and whether or not the SNe are flux-averaged.
We will also examine the impact on the model-independent estimate of the distance-redshift
relation from the SNe.

We describe our method in Sec.II, present our results in Sec.III, and conclude in Sec.IV.

\section{Method}
\label{sec:method}

The comoving distance to an object at redshift $z$ is given by:
\ba
\label{eq:r(z)}
 & &r(z)=cH_0^{-1}\, |\Omega_k|^{-1/2} {\rm sinn}[|\Omega_k|^{1/2}\, \Gamma(z)],\\
 & &\Gamma(z)=\int_0^z\frac{dz'}{E(z')}, \hskip 1cm E(z)=H(z)/H_0 \nonumber
\ea
where ${\rm sinn}(x)=\sin(x)$, $x$, $\sinh(x)$ for $\Omega_k<0$, $\Omega_k=0$, and $\Omega_k>0$ respectively.
The Hubble parameter is given by
\be
 H^2(z)= H_0^2 \left[ \Omega_m (1+z)^3 + \Omega_k (1+z)^2 + \Omega_X X(z) \right],
\ee
where $\Omega_m + \Omega_k + \Omega_X=1$, and the dark energy density function $X(z)$ is defined as
\be
X(z) \equiv \frac{\rho_X(z)}{\rho_X(0)}.
\ee
We have omitted the radiation term, which is negligible at redshifts relevant in this work.

To perform the likelihood analysis of the parameters, we consider two cosmological models:
(1) Fixed cosmology background. Per C11, we consider a simplest $\Lambda$CDM model with $\Omega_m=0.26$.
(2) The comoving distance as a free function of $z$ parametrized by the cubic spline interpolation
of its assumed values at a set of $z$ values.
Following \cite{Wang2009}, we parametrize a scaled comoving distance
\be
r_p(z) \equiv \frac{H_0 r(z)}{cz} (1+z)^{0.41}
\label{eq:rp_sn}
\ee
by its values at $z_i=0.14i$, $i=0,1,2,...,10$, which is a 10 bins $r_p(z)$ model.
Note that $z_0=0$, and $r_p(z_0)=1$.
We also consider a 9 bins $r_p(z)$ model with $z_i=0.14i$, $i=1,2,...,8$, and $z_9=1.4$.
(Note that $z = 1.4$ is the highest redshift of SNe Ia in the SNLS3 dataset).
The $r_p(z)$ at arbitrary $z$ is given by cubic spline interpolation, thus {\it no} assumptions are made about cosmological models.
Therefore, the $\{r_p(z_i)\}$ provide model-independent distance measurements from SNe Ia.

SN Ia data give measurements of the luminosity distance $d_L(z)$ through that of the distance modulus of each SN:
\be
\label{eq:m-M}
\mu_0 \equiv m-M= 5 \log\left[\frac{d_L(z)}{\mathrm{Mpc}}\right]+25,
\ee
where $m$ and $M$ represent the apparent and absolute magnitude of a SN.
The luminosity distance $d_L(z)=(1+z)\, r(z)$, with the comoving distance $r(z)$ given by Eq.(\ref{eq:r(z)}).

Here we use the SNLS3 dataset.
As mentioned above, based on different light-curve fitters,
three SN sets of SNLS3 are given, including
``SALT2'', which consists of 473 SNe Ia; ``SiFTO'', which consists of 468 SNe Ia; and ``combined'', which consist of 472 SNe Ia.
Per C11, here we mainly use the ``combined'' set,
but both ``SALT2'' and ``SiFTO'' sets are also used for comparative studies.

One of the main purposes of our work is to explore the possible evolution of $\alpha$ and $\beta$.
To do this, here we consider three functional forms:
(1) linear case: $\alpha(z) = \alpha_{0} + \alpha_{1} z$ and $\beta(z) = \beta_{0} + \beta_{1} z$
(2) quadratic case: $\alpha(z) = \alpha_{0} + \alpha_{1} z + \alpha_{2} z^2$ and $\beta(z) = \beta_{0} + \beta_{1} z + \beta_{2} z^2$
(3) step function case: per C11, the redshift range [0, 1] is evenly divided into 9 bins,
with both $\alpha$ and $\beta$ constant within each bin.
For comparison, the constant $\alpha$ and $\beta$ case is also considered.
Now, the predicted magnitude of a SN becomes,
\be
m_{\rm mod}=5 \log_{10}{\cal D}_L(z|\mbox{\bf p})
- \alpha(z) (s-1) +\beta(z) {\cal C} + {\cal M},
\ee
where ${\cal D}_L(z|\mbox{\bf p})$ is the luminosity distance
multiplied by $H_0$
for a given set of cosmological parameters $\{ {\bf q} \}$,
$s$ is the stretch measure of the SN light curve shape, and
${\cal C}$ is the color measure for the SN.
${\cal M}$ is a nuisance parameter representing some combination
of the absolute magnitude of a fiducial SN Ia, $M$, and the
Hubble constant, $H_0$.
Since the time dilation part of the observed luminosity distance depends
on the total redshift $z_{\rm hel}$ (special relativistic plus cosmological),
we have
\be
{\cal D}_L(z|\mbox{\bf s})\equiv c^{-1}H_0 (1+z_{\rm hel}) r(z|\mbox{\bf s}),
\ee
where $z$ and $z_{\rm hel}$ are the CMB restframe and heliocentric redshifts of the SN.

For a set of $N$ SN with correlated errors, we have \cite{SNLS3Conley}
\be
\label{eq:chi2_SN}
\chi^2=\Delta \mbox{\bf m}^T \cdot \mbox{\bf C}^{-1} \cdot \Delta\mbox{\bf m}
\ee
where $\Delta m \equiv m_B-m_{\rm mod}$ is a vector with $N$ components,
$m_B$ is the rest-frame peak B-band magnitude of the SN,
and $\mbox{\bf C}$ is the $N\times N$ covariance matrix of the SN.
Note that $\Delta m$ is equivalent to $\Delta \mu_0$, since
\be
\Delta m \equiv m_B-m_{\rm mod}
= \left[m_B+\alpha(z) (s-1) -\beta(z) {\cal C}\right] - {\cal M}.
\ee

The total covariance matrix is \cite{SNLS3Conley}
\be
\mbox{\bf C}=\mbox{\bf D}_{\rm stat}+\mbox{\bf C}_{\rm stat}
+\mbox{\bf C}_{\rm sys},
\ee
with the diagonal part of the statistical uncertainty given by \cite{SNLS3Conley}
\ba
\mbox{\bf D}_{\rm stat,ii}&=&\sigma^2_{m_B,i}+\sigma^2_{\rm int}
+ \sigma^2_{\rm lensing}+ \sigma^2_{{\rm host}\,{\rm correction}} \nonumber\\
&& + \left[\frac{5(1+z_i)}{z_i(1+z_i/2)\ln 10}\right]^2 \sigma^2_{z,i} \nonumber\\
&& +\alpha(z_i)^2 \sigma^2_{s,i}+\beta(z_i)^2 \sigma^2_{{\cal C},i} \nonumber\\
&& + 2 \alpha(z_i) C_{m_B s,i} - 2 \beta(z_i) C_{m_B {\cal C},i} \nonumber\\
&& -2\alpha(z_i)\beta(z_i) C_{s {\cal C},i},
\ea
where $C_{m_B s,i}$, $C_{m_B {\cal C},i}$, and $C_{s {\cal C},i}$
are the covariances between $m_B$, $s$, and ${\cal C}$ for the $i$-th SN,
$\alpha_i=\alpha(z_i)$ and $\beta_i=\beta(z_i)$ are the values of $\alpha$ and $\beta$ for the $i$-th SN.
Note also that $\sigma^2_{z,i}$ includes a peculiar velocity residual of 0.0005 (i.e., 150$\,$km/s) added in quadrature \cite{SNLS3Conley}.
Per C11, here we fix $\sigma_{int}$ to ensure that $\chi^2/dof=1$. 
Varying $\sigma_{int}$ could have a significant impact on parameter estimation, see \cite{Kim2011} for details. 
We will study the effect of varying $\sigma_{int}$ in future work.

We define $\mbox{\bf V} \equiv \mbox{\bf C}_{\rm stat} + \mbox{\bf C}_{\rm sys}$,
where $\mbox{\bf C}_{\rm stat}$ and $\mbox{\bf C}_{\rm sys}$ are the statistical and systematic covariance matrices, respectively.
After treating $\alpha$ and $\beta$ as functions of $z$,
$\mbox{\bf V}$ is given in the form:
\ba
\mbox{\bf V}_{ij}&=&V_{0,ij}+\alpha_i\alpha_j V_{a,ij} + \beta_i\beta_j V_{b,ij} \nonumber\\
&& +\alpha_j V_{0a,ij} +\alpha_i V_{0a,ji} \nonumber\\
&& -\beta_j V_{0b,ij} -\beta_i V_{0b,ji} \nonumber\\
&& -\alpha_i\beta_j V_{ab,ij} - \alpha_j\beta_i V_{ab,ji}.
\ea
It must be stressed that, while $V_0$, $V_{a}$, $V_{b}$, are the same as the ``normal'' covariance matrices
given by the SNLS data archive, $V_{0a}$, $V_{0b}$, and $V_{ab}$ are {\it not} the same as the ones given there.
This is because the original matrices of SNLS3 are produced by assuming that both $\alpha$ and $\beta$ are constant.
We have used the $V_{0a}$, $V_{0b}$, and $V_{ab}$ matrices for the ``combined'', ``SALT2'', and ``SiFTO'' sets
that are applicable when varying $\alpha(z)$ and $\beta(z)$ (A.~Conley, private communication, 2013).

We also use flux averaging of SNe Ia, which was proposed to reduce the effect of weak lensing of SNe Ia on cosmological
parameter estimation \cite{Wang2000}.
Wang \& Mukherjee (2004) \cite{WangPia04} and Wang (2005) \cite{Wang2005} developed a consistent framework for flux-averaging SNe Ia.
Wang, Chuang, \& Mukherjee (2012) \cite{WangChuPia2012} gave the recipe for applying flux averaging method to the SNLS3 data.
Here we just use the method of \cite{WangChuPia2012}.
We refer the reader to Wang, Chuang, \& Mukherjee (2012) \cite{WangChuPia2012} for more details (see also \cite{WangWang2013}).

For the SN Ia samples used in this work, we flux-averaged the SN with $dz=0.04$,
with two different values for the redshift cut off:
$z_{cut} = 0.04$ (flux-averaging only SNe Ia at $z\ge 0.04$),
and $z_{cut} = 0$ (flux-averaging all SNe Ia).
Note that the choice of $z_{cut}=0.04$ is not special; we found
similar results for other choices of $z_{cut}$ (e.g., $z_{cut}=0.2$).
We have omitted the results of other choices of $z_{cut}$ for simplicity and clarity.
Our SN flux-averaging code is available at http://www.nhn.ou.edu/$\sim$wang/SNcode/.

\section{Results}

We perform a Markov Chain Monte Carlo (MCMC) likelihood analysis \cite{COSMOMC}
to obtain ${\cal O}$($10^6$) samples for each set of results presented in this paper.
We assum flat priors for all the parameters, and allow ranges of the parameters wide enough
such that further increasing the allowed ranges has no impact on the results.
The chains typically have worst e-values (the variance(mean)/mean(variance) of 1/2 chains) much smaller than 0.05, indicating convergence.

In the following, we will discuss the results for the cases without and with flux averaging, respectively.

\subsection{No flux averaging cases}

\begin{figure}
\includegraphics[scale=0.4, angle=0]{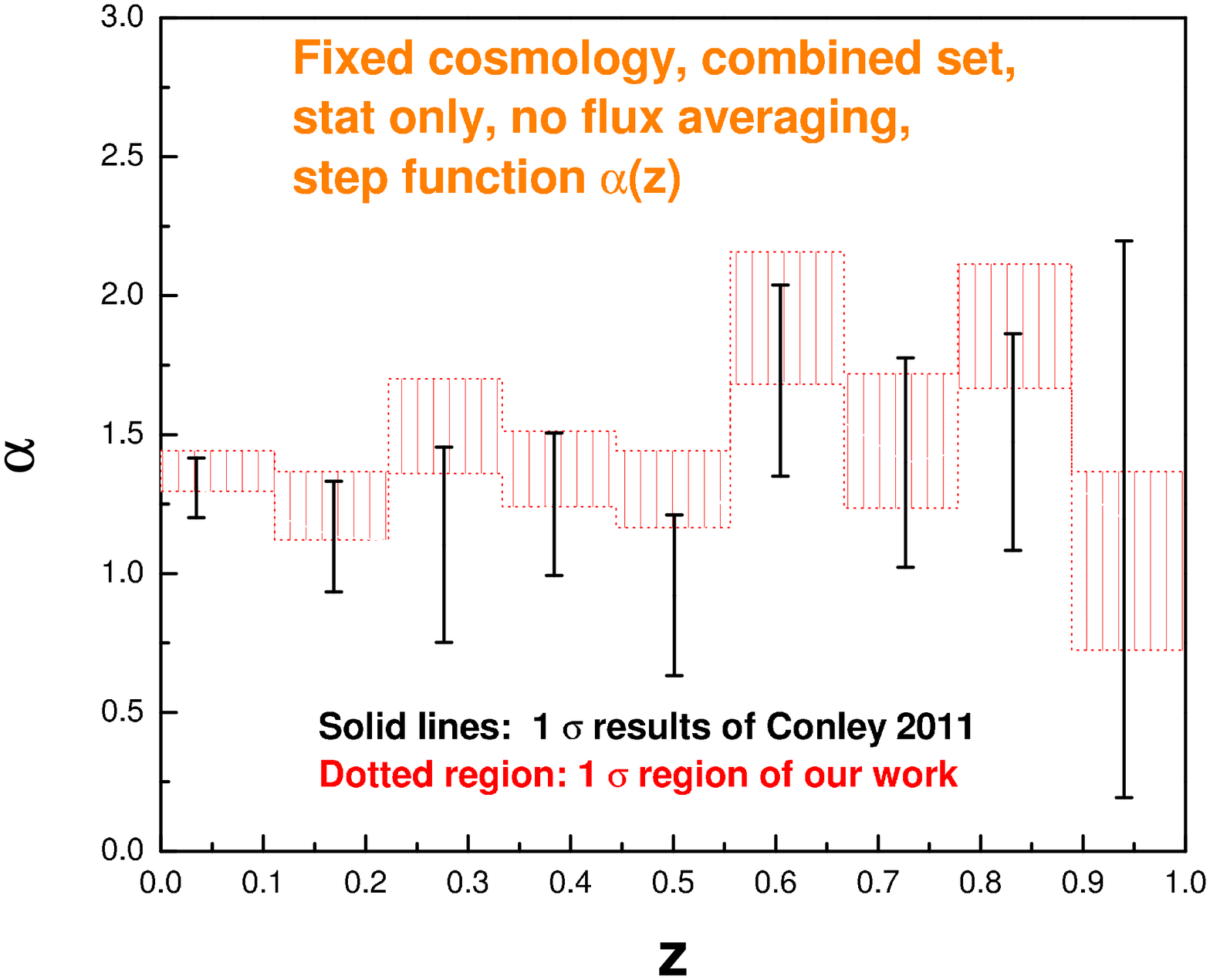}
\includegraphics[scale=0.4, angle=0]{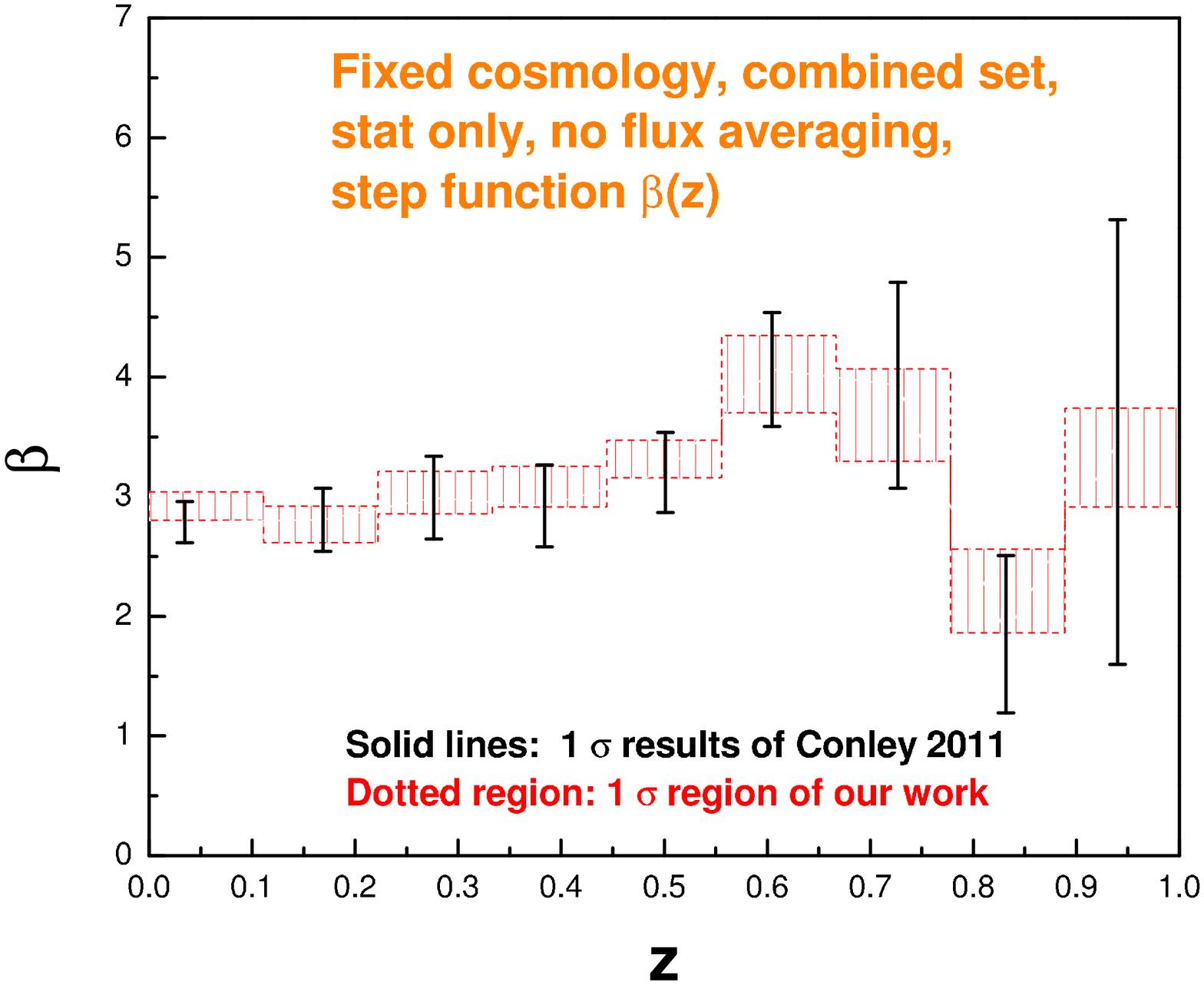}
\caption{\label{fig1}\footnotesize%
The step function $\alpha(z)$ (top panel) and $\beta(z)$ (bottom panel)
from the combined set, including statistical errors only.
To make a direct comparison, the corresponding result of C11 is also shown.
A fixed cosmological model is assumed.}
\end{figure}

\begin{figure}
\includegraphics[scale=0.4, angle=0]{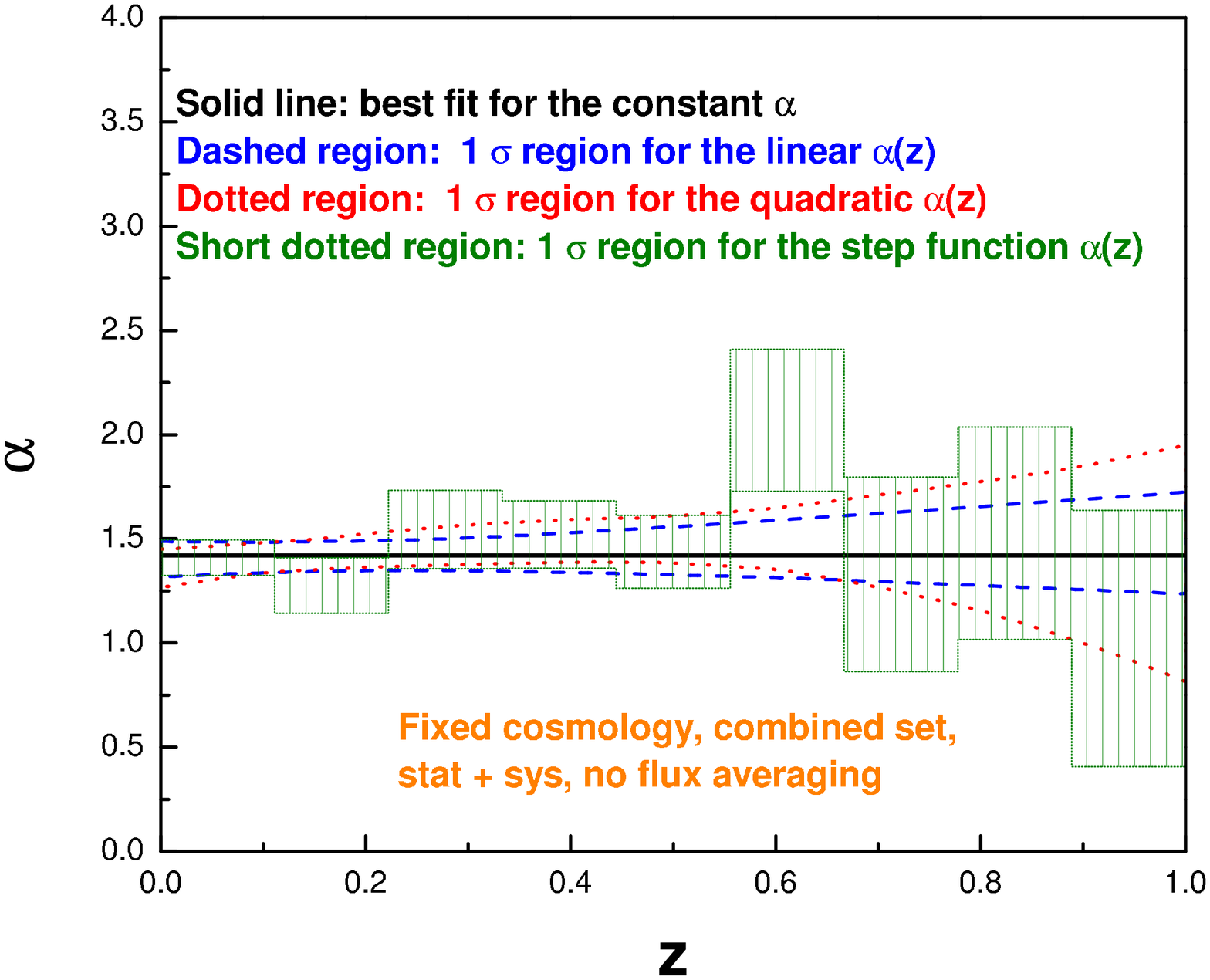}
\includegraphics[scale=0.4, angle=0]{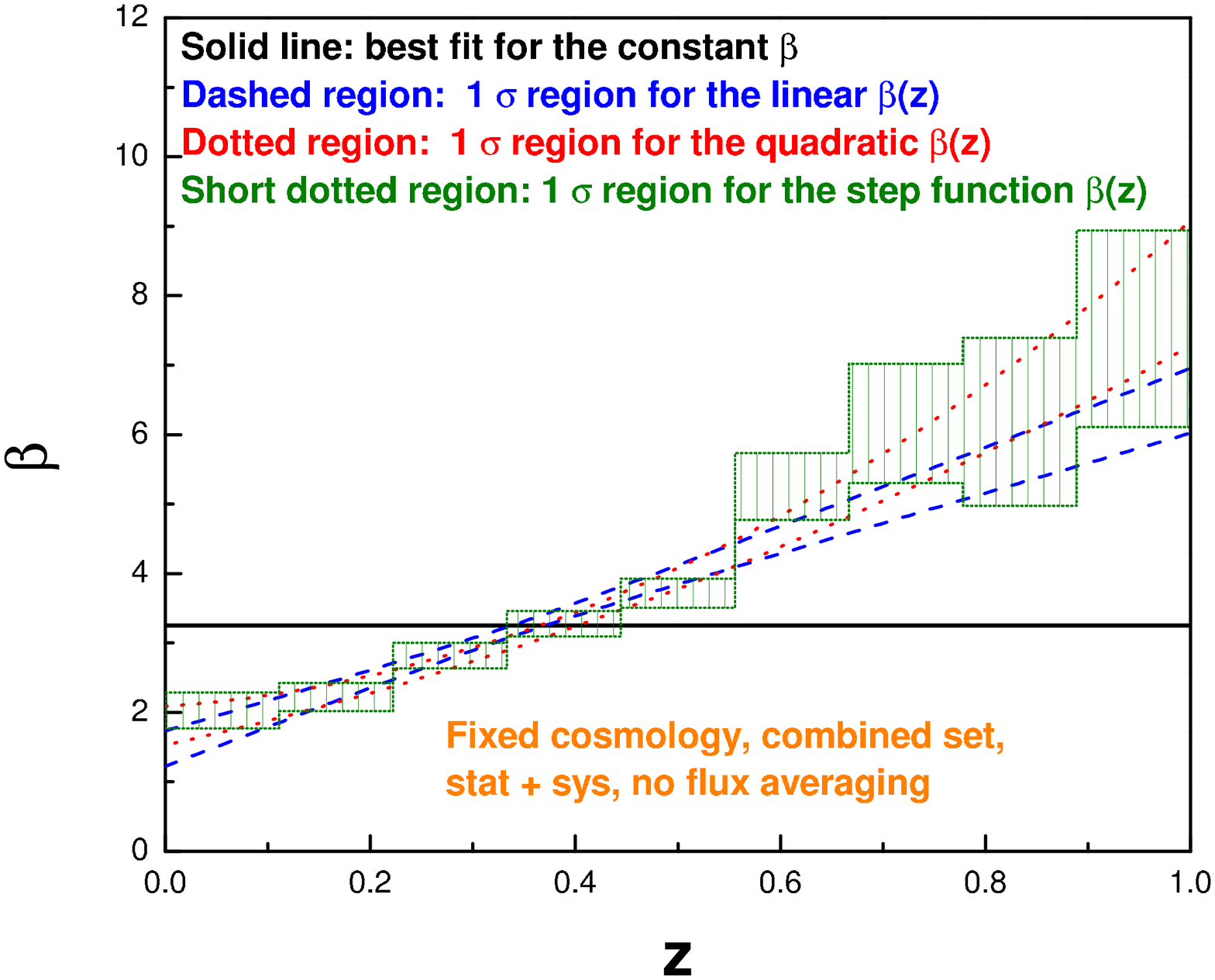}
\caption{\label{fig2}\footnotesize%
The evolution of $\alpha(z)$ (top panel) and $\beta(z)$ (bottom panel)
for the combined SNLS3 set, including both statistical and systematics errors.
A fixed cosmological model is assumed.}
\end{figure}

\begin{figure}
\psfig{file=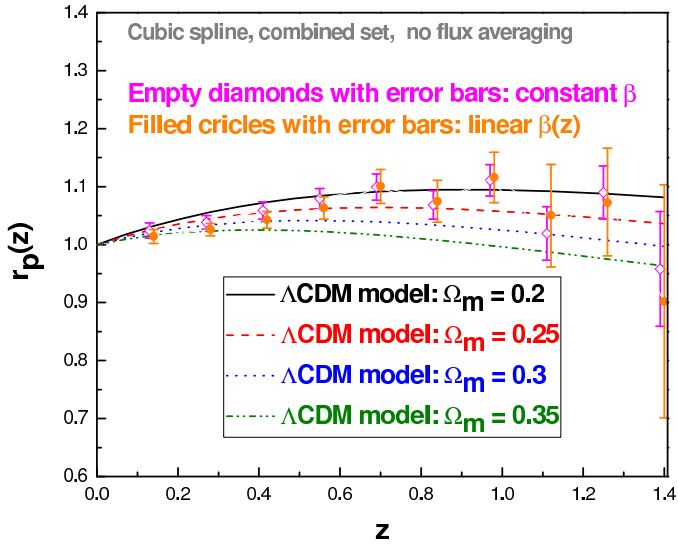,width=3.5in}\\
\caption{\label{fig3}\footnotesize%
The 10 bins $r_p(z_i)$ results given by constant $\beta$ and linear $\beta(z)$,
where the combined set with both statistical and systematics errors are used.
The $r_p(z)$ results of the $\Lambda$CDM models with different $\Omega_m$ are also shown for comparison.
}
\end{figure}

\begin{figure}
\psfig{file=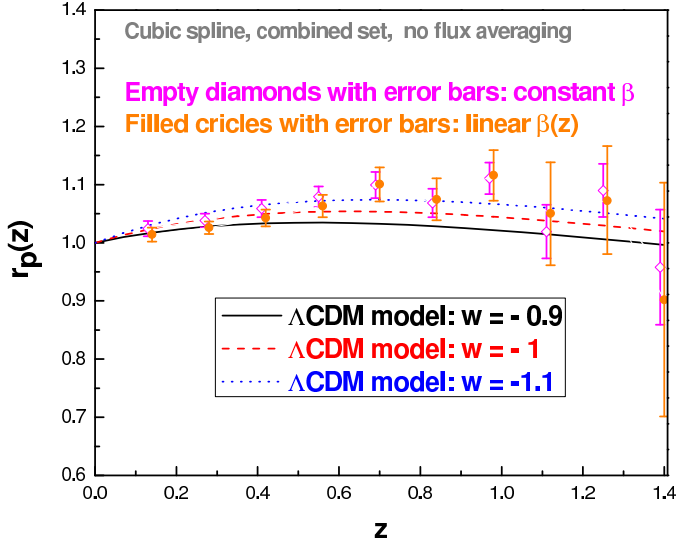,width=3.5in}\\
\caption{\label{fig4}\footnotesize%
The same 10 bins $r_p(z_i)$ results given by constant $\beta$ and linear $\beta(z)$.
The $r_p$ results of XCDM models (fixing $\Omega_m=0.26$) with different $w$ are also plotted for comparison.
}
\end{figure}

\begin{figure}
\includegraphics[scale=0.4, angle=0]{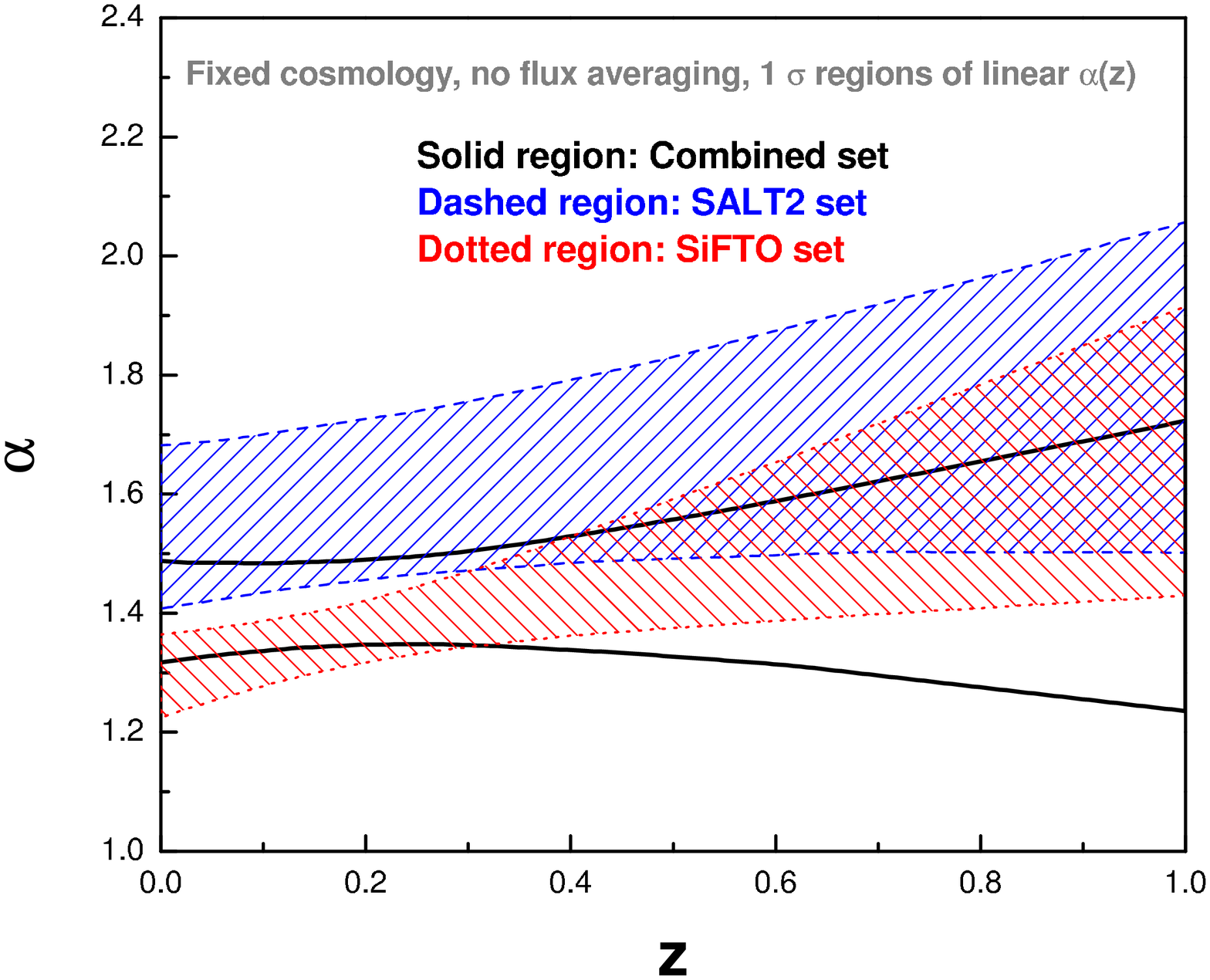}
\includegraphics[scale=0.4, angle=0]{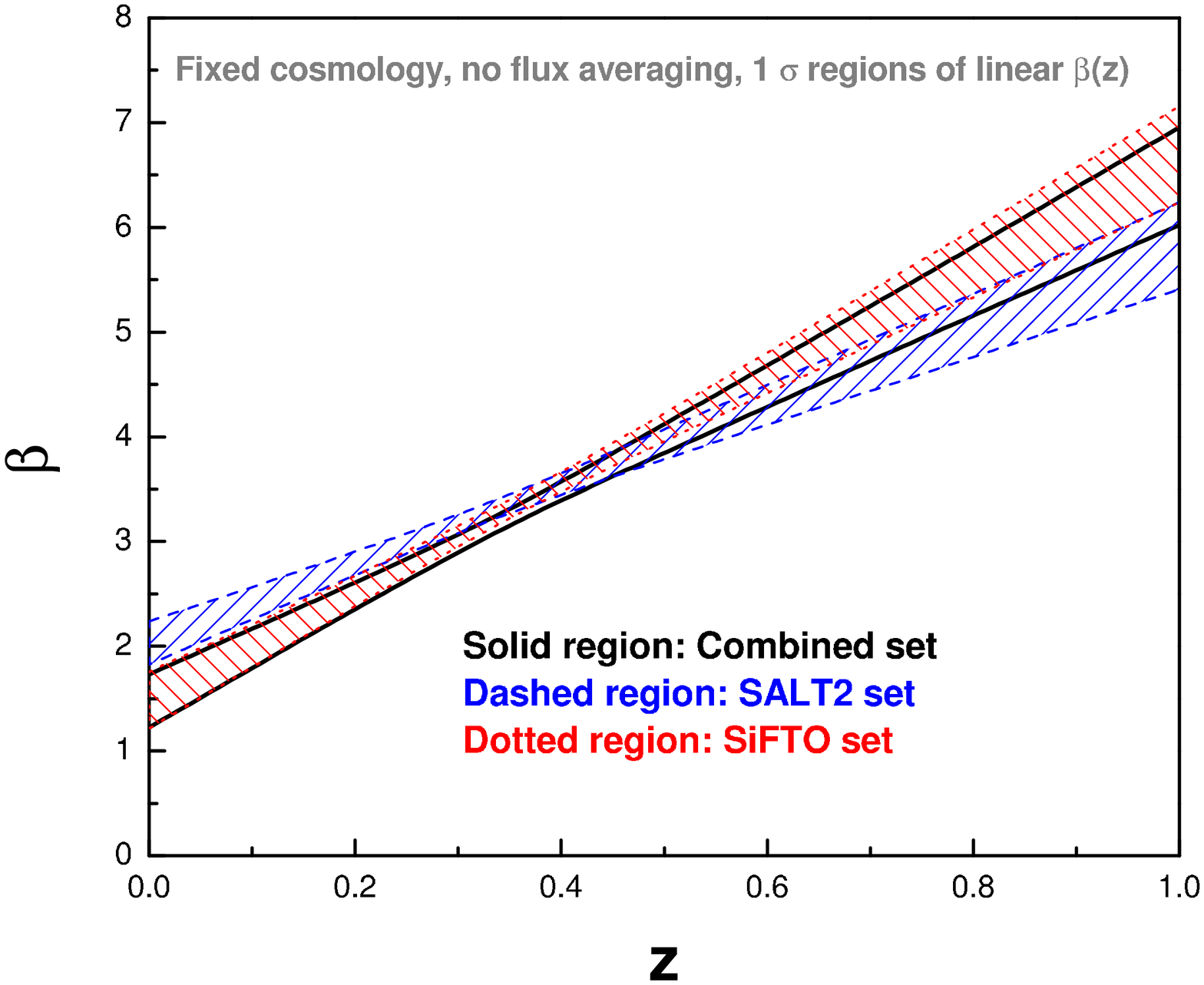}
\caption{\label{fig5}\footnotesize%
A comparison of the evolutions of linear $\alpha(z)$ (top panel) and $\beta(z)$ (bottom panel) from different SN sets,
where both the statistical and systematics errors are used.
Here the fixed cosmology background is adopted.}
\end{figure}

In this subsection, we show the no-flux-averaging results in Figs.\ref{fig1}-\ref{fig5}.

We start with a comparison with Figure 14 of C11.
In Fig.\ref{fig1}, we show the step function $\alpha(z)$ (top panel) and $\beta(z)$ (bottom panel)
from the combined set, including statistical errors only.
To make a direct comparison, the corresponding result of C11 is also shown.
Since the result of C11 is obtained using a $\chi^2$ type analysis with a Hessian,
while our result is obtained using an MCMC analysis, there are small differences between them,
but the trends of $\alpha(z)$ and $\beta(z)$ are essentially the same.
This means that these two results are consistent with each other.
As seen in this figure, considering the statistical errors of combined set only,
there is no sufficient evidence for the evolution of $\alpha$ and $\beta$.

Now, we study the effect of including systematics errors of SNLS3,
in the context of the possible time evolution of $\alpha$ and $\beta$.
The corresponding results are shown in Table \ref{table1} and Fig.\ref{fig2}.
Here we consider three functional forms of $\alpha(z)$ and $\beta(z)$: linear, quadratic, and a step function.
All these three cases give the same results:
$\alpha$ is still consistent with a constant, but $\beta$ now rapidly increases with $z$.
In other words, for the combined set,
there is strong evidence (more than 3 $\sigma$) for the deviation of $\beta$ from a constant.
This is similar to the findings for SDSS \cite{SDSS} and Union2 \cite{Mohlabeng} SNe Ia datasets.
Our results here differ from that of C11 for $\beta$; C11 found that the inclusion of systematic errors
does {\it not} lead to significant redshift dependence of $\beta$.

\begin{table} \caption{Fitting results for various $\alpha(z)$ and $\beta(z)$ models (including systematic errors). }
\begin{center}
\label{table1}
\begin{tabular}{cccc}
  \hline
Model  ~~~  &   Constant  ~~~     &    Linear   ~~~      &   Quadratic    ~~~  \\
  \hline
  $\alpha_{0}$  ~~~  & $1.421^{+0.076}_{-0.077}$   ~~~   &   $1.400^{+0.088}_{-0.083}$   ~~~  &  $1.337^{+0.115}_{-0.072}$ ~~~  \\
    \hline
  $\alpha_{1}$  ~~~  &             ~~~                   &   $0.083^{+0.274}_{-0.285}$   ~~~  &  $0.731^{+0.514}_{-0.895}$  ~~~  \\
    \hline
  $\alpha_{2}$  ~~~  &             ~~~                   &             ~~~                    &  $-0.780^{+1.378}_{-0.816}$  ~~~  \\
    \hline
  $\beta_{0}$   ~~~  & $3.251^{+0.077}_{-0.077}$  ~~~    &   $1.480^{+0.251}_{-0.256}$   ~~~  &  $1.829^{+0.259}_{-0.297}$  ~~~  \\
  \hline
  $\beta_{1}$   ~~~  &             ~~~                   &   $4.998^{+0.723}_{-0.679}$   ~~~  &  $2.081^{+1.456}_{-1.233}$  ~~~  \\
  \hline
  $\beta_{2}$   ~~~  &             ~~~                   &             ~~~                    &  $4.194^{+1.767}_{-1.772}$    ~~~  \\
  \hline
\end{tabular}
\end{center}
\end{table}

Next, we discuss the effect of varying $\beta$ on cosmological parameter estimation.
As shown in Fig.\ref{fig2}, the evolution of linear, quadratic, and a step function $\alpha(z)$ are very similar; the same is true for $\beta(z)$.
In addition, by studying the linear $\alpha(z)$ and $\beta(z)$ case,
we find that varying $\alpha$ has no significant impact on parameter estimation. 
Thus we will assume a constant $\alpha$ and a linear $\beta(z)$ from now on.
For the constant $\beta$ and the linear $\beta(z)$ cases, we show the results of 10 bins $r_p(z_i)$ in Fig.\ref{fig3}.
Note that the $r_p(z)$ at arbitrary $z$ is given by cubic spline interpolation,
thus {\it no} assumptions are made about cosmological models.
The $\{r_p(z_i)\}$ thus provide model-independent distance measurements from SNe Ia.
For comparison, we also show the $r_p(z)$ of the $\Lambda$CDM models with different $\Omega_m$.
As seen in this figure,
linear $\beta(z)$ yields smaller $\{r_p(z_i)\}$ results at low redshift,
which are closer to the result of the $\Lambda$CDM model with a larger $\Omega_m$.
In addition, at high redshift,
the error bars of $\{r_p(z_i)\}$ given by linear $\beta(z)$ are larger than those given by constant $\beta$.
This means that treating $\beta$ as a constant during the cosmology-fits
will underestimate the errors of model parameters, if the time dependence of $\beta$ is real.

In Fig.\ref{fig4}, we compare the $r_p(z_i)$ results of Fig.\ref{fig3} with the $r_p(z)$ results of XCDM models with different values
of constant $w$. Based on Fig.\ref{fig3} and Fig.\ref{fig4},
one can see that the time variation of $\beta$ has a significant impact on cosmological parameter estimation.
This shows the importance of considering the evolution of $\beta$ during the cosmology-fits.
Assuming a linear $\beta(z)$ leads to a distance-redshift relation that is closer to a cosmological constant,
compared to that from assuming constant $\beta$.
In addition, the error bars of $\{r_p(z_i)\}$ parameters for a linear $\beta(z)$
are increased by $50\%$ - $100\%$ at high $z$, compared to a constant $\beta$. 

Finally, we study the impact of using different SN lightcurve fitters
on the evolution of linear $\alpha(z)$ and $\beta(z)$.
Using the ``combined'', ``SALT2'', and ``SiFTO'' sets respectively,
we plot the linear $\alpha(z)$ and $\beta(z)$ in Fig.\ref{fig5}.
As seen in this figure,
although there are some differences in details,
the trends of $\alpha(z)$ and $\beta(z)$ given by these three SN sets are same.
Again, we see strong evidence for the redshift-dependence of $\beta$, for all the SN sets of SNLS3.

\subsection{Flux-averaging cases}

\begin{figure}
\psfig{file=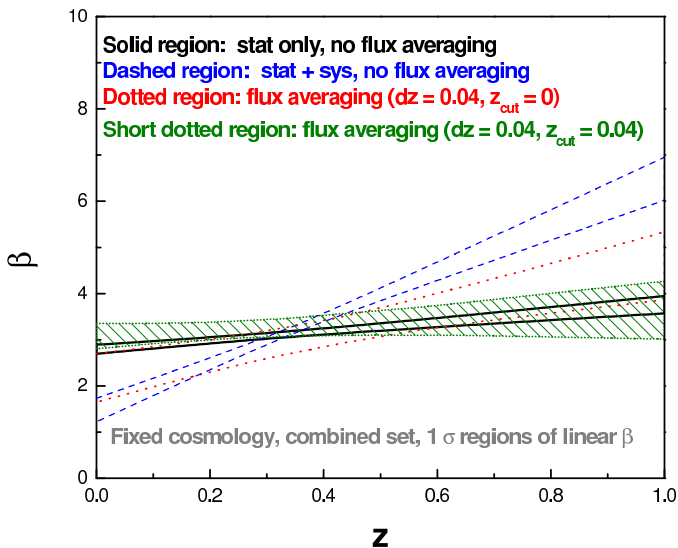,width=3.5in}\\
\caption{\label{fig6}\footnotesize%
The evolutions of linear $\beta(z)$ from the combined SNLS3 set,
assuming a fixed cosmological model.
Three cases are plotted:
1. statistical errors only, no flux-averaging.
2. statistical and systematic errors, no flux-averaging.
3. statistical and systematic errors, flux averaging with $dz=0.04$ and two different choices of $z_{cut}$ (0 and 0.04).
}
\end{figure}

\begin{figure}
\psfig{file=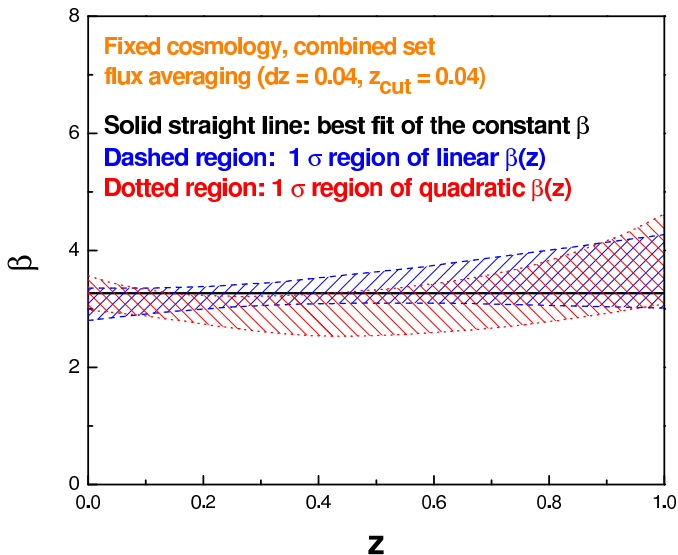,width=3.5in}\\
\caption{\label{fig7}\footnotesize%
The evolution of $\beta(z)$ from the combined SNLS3 set,
including both statistical and systematics errors.
Here we assume a fixed cosmological model, and flux-average SNe Ia with $dz=0.04$ and $z_{cut}=0.04$.
}
\end{figure}

\begin{figure}
\psfig{file=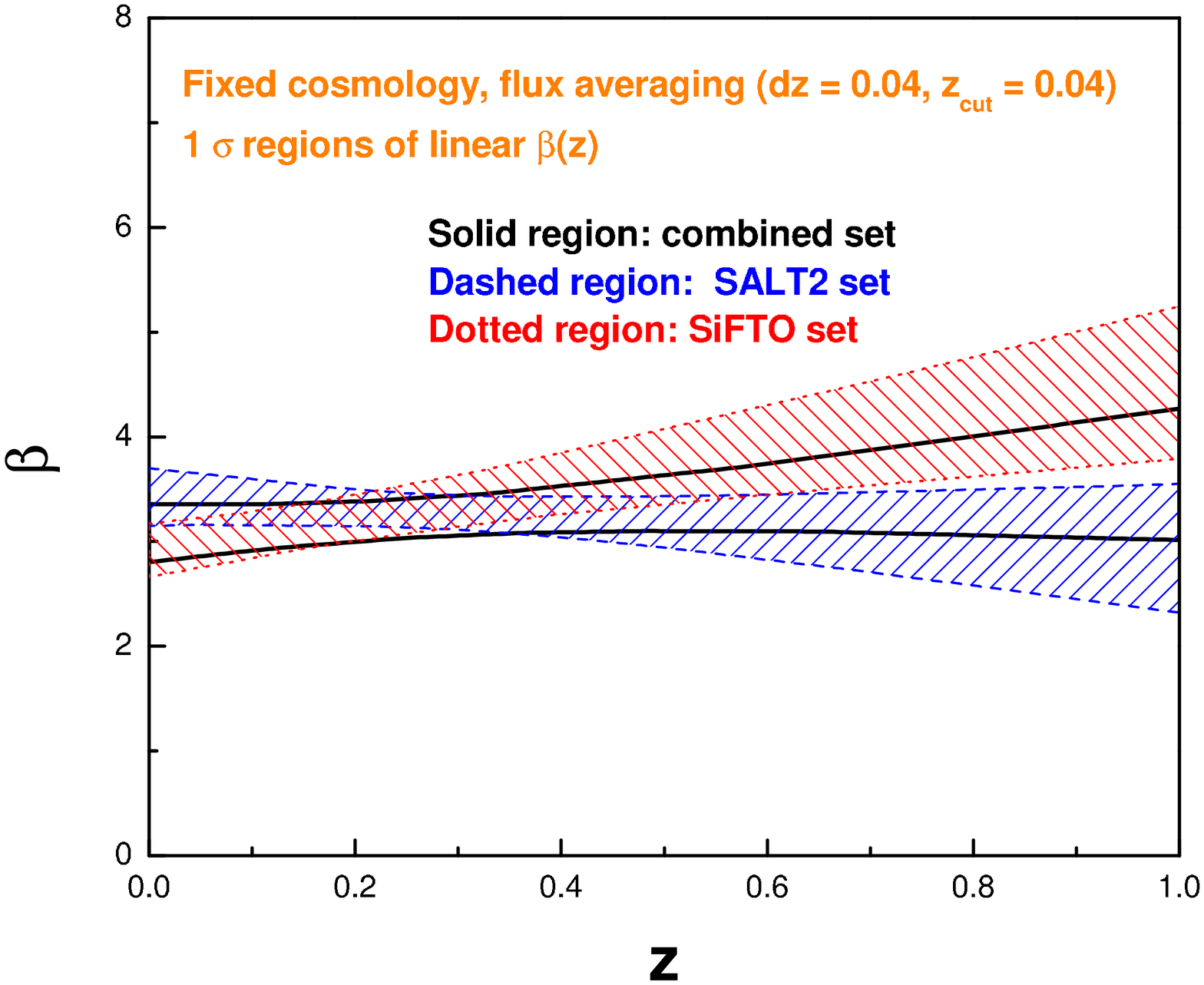,width=3.5in}\\
\psfig{file=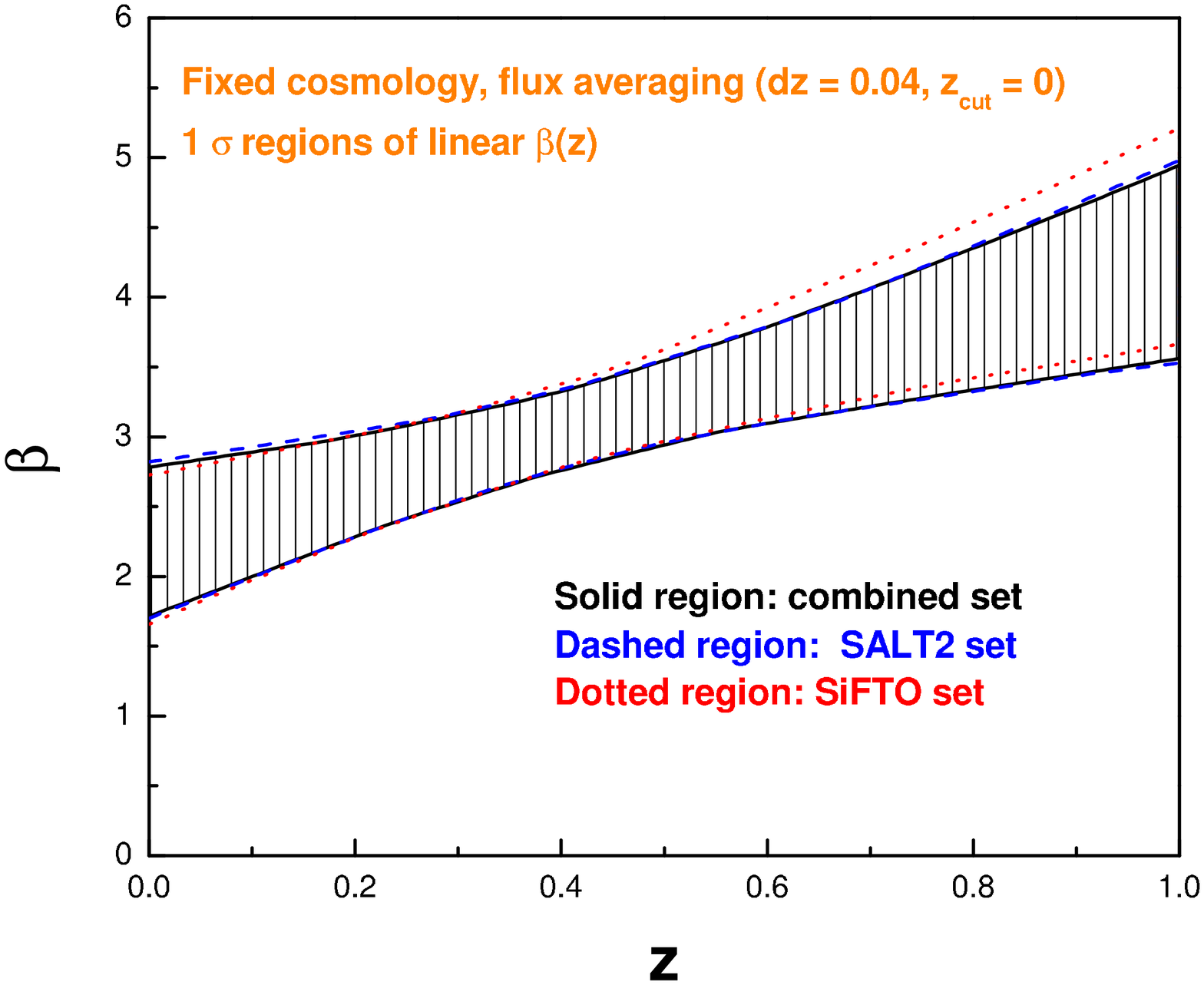,width=3.5in}\\
\caption{\label{fig8}\footnotesize%
A comparison of the linear $\beta(z)$ from different SNLS3 data sets,
including both statistical and systematics errors.
We assume a fixed cosmological model, and flux-average SNe Ia with $dz=0.04$ and $z_{cut}=0.04$ (top panel)
and $z_{cut}=0$ (bottom panel).
}
\end{figure}

\begin{figure}
\psfig{file=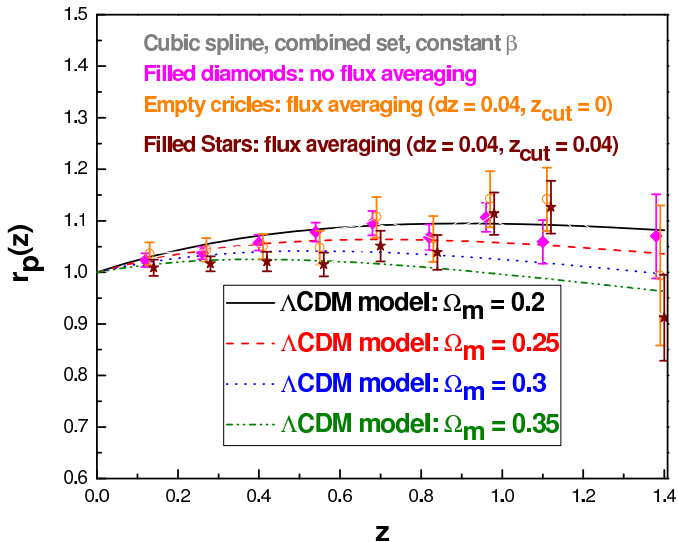,width=3.5in}\\
\caption{\label{fig9}\footnotesize%
A comparison of the 9 bins $r_p(z_i)$ results of constant $\beta$ among the cases without and with flux-averaging,
using the combined SNLS3 data set with both statistical and systematics errors.
Two flux-averaging cases are considered here: $z_{cut}=0$ and $z_{cut}=0.04$.
The $r_p(z)$ results of the $\Lambda$CDM models with different $\Omega_m$ are shown for comparison.
}
\end{figure}

\begin{figure}
\psfig{file=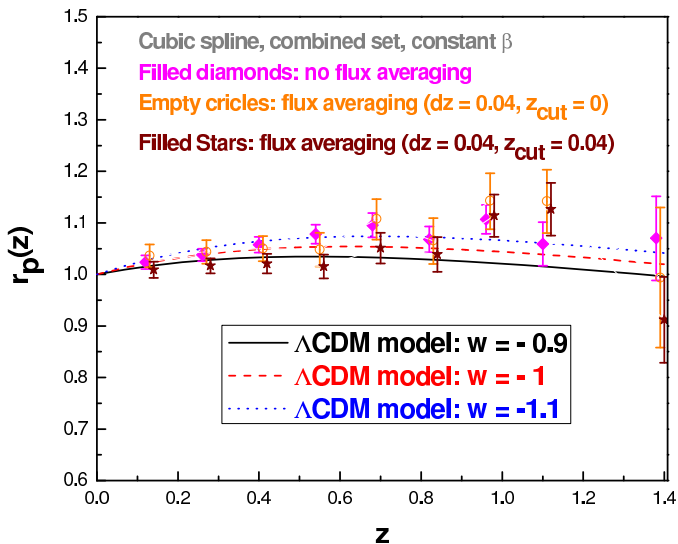,width=3.5in}\\
\caption{\label{fig10}\footnotesize%
The same 9 bins $r_p(z_i)$ results given by constant $\beta$.
The $r_p(z)$ results of XCDM models (fixing $\Omega_m=0.26$) with different $w$ are plotted for comparison.
}
\end{figure}

In this subsection, we study the impact of flux-averaging SNe Ia, in comparison with the no-flux-averaging cases.
Since $\alpha$ is always consistent with a constant, here we only present the results of $\beta$.
The corresponding results are shown in Figs.\ref{fig6}-\ref{fig10}.

Fig.\ref{fig6} shows the evolution of linear $\beta(z)$ from the combined SNLS3 set,
where a fixed cosmological model is assumed.
We consider three cases:
1. statistical errors only, no flux-averaging.
2. statistical + systematic errors, no flux-averaging.
3. statistical + systematic errors, flux-averaging with $dz=0.04$ and two different choices of $z_{cut}$ (0 and 0.04).
As shown in last subsection, for the no-flux-averaging case, adding systematic errors will cause $\beta$ to vary with $z$.
Now, we can see that, for the statistical + systematic errors case, applying flux-averaging of SNe Ia will significantly
decrease the time-evolution of $\beta$: it becomes marginal for $z_{cut}=0$, and is consistent with zero for $z_{cut}=0.04$.

In Fig.\ref{fig7}, we plot constant $\beta$, linear $\beta(z)$, and quadratic $\beta(z)$ from the combined SNLS3 set,
including both statistical and systematics errors.
We assumed a fixed cosmological model, and flux-average SNe Ia with $dz=0.04$ and $z_{cut}=0.04$.
This figure shows that for both linear $\beta(z)$ and quadratic $\beta(z)$ cases,
after flux-averaging the SNe Ia, $\beta$ is consistent with a constant.
This shows that the results shown in Fig.\ref{fig6} are not sensitive to the assumed functional forms of $\beta(z)$.

In Fig.\ref{fig8}, we show the impact of different SN lightcurve fitters on the evolution of linear $\beta(z)$,
including both statistical and systematics errors.
We assume a fixed cosmological model, and flux-average SNe Ia with $dz=0.04$ and $z_{cut}=0.04$ (top panel)
and $z_{cut}=0$ (bottom panel).
Again, we see that although there are some differences in details,
the trends of $\beta(z)$ given by three SNLS3 data sets (``Combined'', ``SALT2'', and ``SiFTO'') are the same,
i.e. consistent with a constant.
This shows that the result of Fig.\ref{fig6} is also independent of the SN Ia lightcurve fitter used.

Finally, we discuss the effect of using flux-averaging on cosmological parameter estimation.
Since flux-averaging reduces the evolution of $\beta$ to be marginal or zero,
here we just consider the constant $\beta$ case.
For the cases without and with flux-averaging, we show the results of 9 bins $r_p(z_i)$ in Fig.\ref{fig9}.
Two flux-averaging methods are considered here: $z_{cut}=0$ and $z_{cut}=0.04$.
For comparison, we also plot the $r_p(z)$ of the $\Lambda$CDM models with different $\Omega_m$.
It is interesting to note that flux averaging all SNe ($z_{cut}=0$) or just those at $z\ge 0.04$
($z_{cut}=0.04$) give noticeably different results:
using $z_{cut}=0$ gives $r_p(z_i)$ closer to the no-flux-averaging case, while
using $z_{cut}=0.04$ gives $r_p(z_i)$ that are significantly different from both.

Fig.\ref{fig10} compares the $r_p(z_i)$ results of Fig.\ref{fig9} with the $r_p(z)$ results of XCDM models with
different values of constant $w$.
The results of Fig.\ref{fig9} and Fig.\ref{fig10} show that
it is important to consider the flux-averaging of SNe Ia during the cosmology-fits.

\section{Discussion and Summary}

We have explored the systematic uncertainties of the SNLS3 dataset by treating $\alpha$ and $\beta$ as functions of $z$.
To do this, we have considered three functional forms for $\alpha$ and $\beta$:
(1) linear case: $\alpha(z) = \alpha_{0} + \alpha_{1} z$ and $\beta(z) = \beta_{0} + \beta_{1} z$
(2) quadratic case: $\alpha(z) = \alpha_{0} + \alpha_{1} z + \alpha_{2} z^2$ and $\beta(z) = \beta_{0} + \beta_{1} z + \beta_{2} z^2$
(3) step function case: the redshift range of [0,1] is evenly divided into 9 bins, with both $\alpha$ and $\beta$ constant within each bin.

To perform the cosmology-fits, we have adopted two cosmological models:
one is a fixed cosmological model, chosen to be the same as that used in C11;
another is the comoving distance obtained by the cubic spline interpolation of
scaled comoving distances, $r_p(z_i)$, measured as free parameters from the data.
These provide model-independent distance measurements from SNe Ia.
For the cubic spline interpolation, we consider two cases:
a 10 bins $r_p(z)$ model with $z_i=0.14i$, $i=1,2,...,10$;
and a 9 bins $r_p(z)$ model with $z_i=0.14i$, $i=1,2,...,8$, and $z_9=1.4$.

We have used the flux averaging of SNe Ia,
which is very useful in reducing the impact of unknown systematic effects of SNe Ia on parameter estimation.
We have flux-averaged the SN with $dz=0.04$.
Two cut off of redshift were chosen: $z_{cut} = 0.04$ and $z_{cut} = 0$.

We find that when SNe Ia are {\it not} flux-averaged, including the systematics errors of SNLS3
leads to strong evidence for the deviation of $\beta$ from a constant (see Fig.\ref{fig2}).
In addition, a time-varying $\beta(z)$ leads to $r_p(z_i)$ estimates closer to that of
a cosmological constant model, compared to that from assuming a constant $\beta$ (see Fig.\ref{fig3} and Fig.\ref{fig4}).

We find when SNe Ia are flux-averaged, the time-evolution of
$\beta$ is significantly reduced (see Fig.\ref{fig6} and Fig.\ref{fig7}).
Flux-averaging all SNe ($z_{cut}=0$) leads to $\beta$ that varies only marginally with time (see Fig.6),
while flux-averaging only the SNe at $z\ge 0.04$ ($z_{cut}=0.04$) leads to $\beta$ being consistent
with a constant.
Surprisingly, these two choices of $z_{cut}$ lead to significantly different distance-redshift relations (see Figs.9 and 10).
This indicates that the unknown systematic biases may originate mostly from low $z$ SNe.
Flux-averaging {\it all} SNe should lead to the least biased results.

We have studied all three SN sets of SNLS3: ``combined'', ``SALT2'' and ``SiFTO''.
We find that the results of $\alpha(z)$ and $\beta(z)$
given by these three SN sets of SNLS3 always have the same trend (see Fig.\ref{fig5} and Fig.\ref{fig8}),
showing that our conclusions are independent of the light-curve fitters of SNe Ia.

Based on the findings on $\alpha(z)$ and $\beta(z)$ from this paper,
we will fully explore the impact of flux-averaging SNe Ia and assumptions on
$\beta(z)$ on dark energy constraints in a companion paper.

It is likely that the apparent evolution of $\beta$ with $z$ for SNe without flux-averaging
is a consequence of unknown systematic effects; e.g., Kessler et al. (2013) \cite{Kessler13} found that
the adoption of the incorrect color model leads to a systematic bias in the derived distance
modulus, which is equivalent to an artificial change in $\beta$.

Flux-averaging reduces the impact of unknow systematic effects by averaging
them within each redshift bin. 
Wang \& Mukherjee (2004) \cite{WangPia04} showed that flux-averaging SNe leads to less biased results by increasing
the likelihood of the true model; see Sec.2.1 of their paper for the details of their analysis.
Although this was done in the context of gravitational lensing of SNe, the same arguments hold
for any unknown systematic effect that averages to zero (when a sufficiently large number of SNe 
are available) within a given redshift bin.

Our understanding of the systematic uncertainties
of SNe Ia will improve as larger and more uniform sets of SNe become available from
future surveys \cite{Wang2000a,jedi,wfirst}.

\bigskip

{\bf Acknowledgments}
We are grateful to Alex Conley for providing us with the SNLS3 covariance matrices that
allow redshift-dependent $\alpha$ and $\beta$, and for very helpful discussions.
We acknowledge the use of CosmoMC.
This work is supported in part by DOE grant DE-FG02-04ER41305.

\end{document}